\documentclass{article}
\usepackage{comment}
\usepackage{bbm}
\usepackage{comment}
\usepackage{adjustbox}
\usepackage[utf8]{inputenc} 
\usepackage[T1]{fontenc}    
\usepackage{hyperref}       
\usepackage{url}            
\usepackage{amsmath}       
\usepackage{enumerate}
\usepackage{lmodern}
\usepackage[english]{babel}
\usepackage{amsfonts}       
\usepackage{amsmath}
\usepackage{nicefrac}       
\usepackage{microtype}      
\usepackage{lipsum}
\usepackage[table,xcdraw]{xcolor}
\usepackage[sort]{natbib}
\usepackage{fancyhdr}       
\usepackage{graphicx}       
\usepackage{float}
\usepackage{colortbl}
\usepackage{multirow}
\usepackage{wrapfig}
\usepackage{float}
\usepackage[singlelinecheck=false]{caption}
\usepackage{subcaption,siunitx,booktabs}
\usepackage{geometry}
\usepackage{acronym}
\acrodef{PCA}[PCA]{Principal Component Analysis}
\acrodef{AE}[AE]{Autoencoder}
\acrodef{PC}[PC]{Principal Component}
\acrodef{LM}[LM]{Linear Model}
\acrodef{SCRT}[SCRT]{Single-Case Randomization Test}
\acrodef{RT}[RT]{Randomization Test}
\acrodef{AR1}[AR1]{with a first order autoregressive error structure}

\usepackage{titlesec}

\titleformat{\paragraph}
{\normalfont\normalsize\bfseries}{\theparagraph}{1em}{}
\titlespacing*{\paragraph}
{0pt}{3.25ex plus 1ex minus .2ex}{1.5ex plus .2ex}

\definecolor{green}{RGB}{93, 208, 80}
\definecolor{lightorange}{RGB}{255,222,202}

\usepackage{xargs}                      
\usepackage[colorinlistoftodos,prependcaption,textsize=tiny]{todonotes}
\newcommandx{\unsure}[2][1=]{\todo[linecolor=red,backgroundcolor=red!25,bordercolor=red,#1]{#2}}
\newcommandx{\change}[2][1=]{\todo[linecolor=blue,backgroundcolor=blue!25,bordercolor=blue,#1]{#2}}
\newcommandx{\info}[2][1=]{\todo[linecolor=OliveGreen,backgroundcolor=OliveGreen!25,bordercolor=OliveGreen,#1]{#2}}
\newcommandx{\improvement}[2][1=]{\todo[linecolor=Plum,backgroundcolor=Plum!25,bordercolor=Plum,#1]{#2}}
\newcommandx{\thiswillnotshow}[2][1=]{\todo[disable,#1]{#2}}

\title{Multimodal Outcomes in N-of-1 Trials: Combining Unsupervised Learning and Statistical Inference}

\author{ 
    {Juliana Schneider$^{1,*}$, Thomas G\"artner$^{1}$, Stefan Konigorski$^{1,2,*}$}\\
	$^{1}$Hasso Plattner Institute for Digital Engineering, Potsdam, Germany\\
	$^{2}$Icahn School of Medicine at Mount Sinai, New York, NY, USA\\
    $^{*}$ Corresponding authors: 
    \url{juliana.schneider@hpi.de}, \\\url{stefan.konigorski@hpi.de}\\
}

\begin{document}

\maketitle

\begin{abstract}
N-of-1 trials are within-person crossover trials allowing both personalized and population-level inference on the effect of health interventions. 
Using the full potential of modern technologies, multimodal N-of-1 trials can integrate multimedia data for measuring health outcomes. However, methodology required for automated applications in large multimodal trials is not available yet.
Here, we present an unsupervised approach for modeling multimodal N-of-1 trials, bypassing the need for expensive outcome labeling by medical experts. 
First, an autoencoder is trained on the outcome medical images.
Then, the dimensionality of embeddings is reduced by extracting the first principal component, which is finally tested for its association with the treatment. 
Results from imaging simulation studies show high power in detecting a treatment effect while controlling type I error rates. An application to imaging N-of-1 trials of acne severity identifies individual treatment effects and supports that our methodology can enable large clinical multimodal N-of-1 trials.
\end{abstract}

\newpage

\section{Introduction}

With the increasing development and use of digital tools for personalized health interventions and recommendations, reliable methods for analyzing individual-level effects of different types of interventions for different types of health conditions are in demand. Patients and digital health consumers may wish to easily and dependably assert the influence of a treatment, medication, or health intervention, and thereby personalize the treatment.
For individual-level inference on the efficacy or effectiveness of health interventions, N-of-1 trials present the gold standard study design. N-of-1 trials are multi-crossover experimental studies in single participants, where the health intervention of interest and a comparator -- which could be standard routine or an alternative intervention -- are alternated one or multiple times \cite{mirzaHistoryDevelopmentNof12017, kravitzDesignImplementationNof1, niklesEssentialGuideNof12015}. These trials are especially useful when the effect of a treatment can be assumed to vary between individuals so that there is no one-size-fits-all treatment, for example in the presence of comorbidities, but also in rare diseases, chronic diseases or for expensive treatments where large-sample randomized controlled trials are not available. 
Most N-of-1 trials have relied on numeric health outcome data such as patient reported outcomes or expert ratings \cite{konigorskiStudyUPlatformDesigning2022}, with some recent work using wearables \cite{zhouAnalyzingPopulationlevelTrials2024a, dazaEffectsSleepDeprivation2020}. However, as mobile devices become more and more technologically advanced, multimodal outcome data -- such as audio, video, or images -- can also be used to assess health outcomes more fine-grained in higher detail. In addition, individuals can be empowered to conduct their own trials on their mobile phones without the need for an expensive human expert evaluation of the health condition.

To the best of our knowledge, the study by Fu et al. \cite{fuMultimodalNof1Trials2023} presents the first and only multimodal N-of-1 trial where health outcomes have been directly assessed using images or any other multimedia modality. In this pilot study, participants conducted N-of-1 trials alternating between applying and not applying a self-selected acne cream. Skin conditions have shown a high treatment heterogeneity \cite{quereuxProspectiveStudyRisk2006, lehmannAcneTherapyMethodologic2002} so that they present a primary use case for N-of-1 trials. However, we are not aware of any N-of-1 trial on acne besides \cite{fuMultimodalNof1Trials2023}. Also for skin conditions more generally, few N-of-1 trials have been performed. Existing studies used physical markers or rating scales to assess health outcomes of interest, and none of the published studies directly used visual changes  \cite{arnoldPlaceboControlledCrossover2009, grahamParticipatoryCodesignPatientreported2020, koideRandomizedNof1Single2015, linRemovalDiethylhexylPhthalate2017, roustitOnDemandSildenafilTreatment2018}.
For acne, its severity is typically assessed through visual inspection by an expert and involves counting of pimples or lesions as well as overall skin irritation \cite{agnewComprehensiveCritiqueReview2016, hayashiEstablishmentGradingCriteria2008, tanCurrentMeasuresEvaluation2008}. Counting pimples generates numeric information that can easily be established from an image, while assessing the overall skin condition is more subjective. Possible indicators for irritation could be redness of the skin or the structure or size of the pimples. A holistic consideration of all characteristics and changes in the skin, as possible by collecting images, might thus improve the detection of treatment effects on acne, and skin conditions more generally.

Including multimodal health outcomes in N-of-1 trials hence carries large promise, but requires the development of novel methodology. While there exist many studies on deep learning architectures for multimodal data in general, this has only been applied to N-of-1 trials in the pilot study by Fu et al. \cite{fuMultimodalNof1Trials2023}. Standard N-of-1 trials with tabular outcome data can be analyzed using statistical methods and causal inference \cite{kravitzDesignImplementationNof1, konigorskiDigitalNof1Trials2025, piccininniCausalInferenceNof12024c, dazaCausalAnalysisSelftracked2018}. The same statistical methods can be used when health outcomes are assessed using wearables \cite{zhouAnalyzingPopulationlevelTrials2024a, dazaEffectsSleepDeprivation2020}.
However, novel approaches are required if health outcomes are assessed using multimedia data such as audio, video, or images, and require the integration of deep learning models with statistical hypothesis testing applied to the single time series with limited available data.
In the analysis of the multimodal acne pilot N-of-1 trial \cite{fuMultimodalNof1Trials2023}, t-tests following a supervised convolutional neural network (CNN) trained on manually generated labels were applied. While the approach was successful in identifying a treatment effect, the approach required manual labelling of the images, which is not feasible for an automated large-scale deployment of such trials \cite{fuMultimodalNof1Trials2023}.

In this study, we propose a novel approach based on automated unsupervised learning that is able to analyze multimodal N-of-1 trials without the need for expert labelling.
The unsupervised learning uses an autoencoder to first find lower-dimensional representations of the images, applies a further linear reduction to one dimension, and then performs statistical association tests to test for individual treatment effects. By training an unsupervised model, labels for each image do not need to be provided. The analysis is then based on the assumption that the embeddings obtained from the unsupervised learning capture the relevant features in the input images, so that they can be used to identify differences in skin condition between intervention and non-intervention phases. For evaluation of the approach, we first perform a simulation study on simulated images. The results indicate that our proposed approach yields valid empirical type I error rates when there is no treatment effect and it has high empirical power when there is an effect of the simulated skin treatment. Next, we evaluate our approach on the published acne trial \cite{fuMultimodalNof1Trials2023} testing whether a skin cream has an effect on acne. The results confirm that our approach is able to identify individual-level treatment effects, and thus may be used to enable future large-scale multimodal N-of-1 trials.

\section{Results} \label{Results}

\subsection{Overview}

\begin{figure}[!ht]
    \centering
    \includegraphics[width=\textwidth]{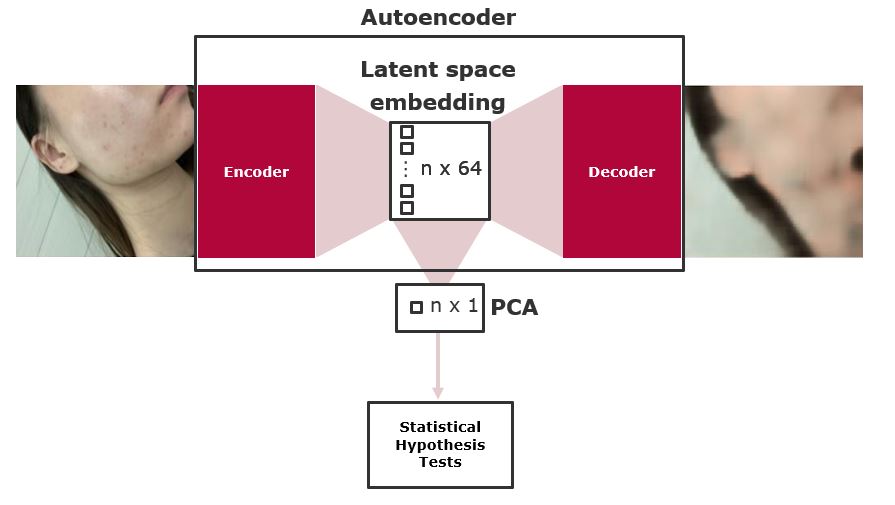}
    \caption{Illustration of the three steps of the proposed approach, containing unsupervised learning, dimensionality reduction and statistical inference. Here, \textit{n} is the number of trial participants the autoencoder is trained on.}
    \label{fig:approach}
\end{figure}

In this section, we first outline our novel proposed approach combining unsupervised learning with statistical hypothesis testing for the analysis of multimodal N-of-1 trials, and then provide results from its evaluation and application. With our approach, we aim to make individual-level inference on whether a given intervention has an effect on a given health outcome that is assessed by images. That is, we are testing whether for a particular individual, the images under treatment are on average different compared to images under no treatment, with respect to some meaningful clinical metrics.
To this aim, we perform the following three steps (see also Figure \ref{fig:approach} for a visualization):
\begin{enumerate}
  \item Obtain an embedding of the images by training an \ac{AE},
  \item Perform a dimensionality reduction of the embeddings via \ac{PCA} and extract the first \ac{PC},
  \item Perform a hypothesis test whether the \ac{PC} scores differ on average between the two treatment conditions.
\end{enumerate} 

In the following, we first describe results from an evaluation of our approach in Monte Carlo simulation studies,
where we insert white circles as acne pimples on images of faces that were extracted randomly from the ACNE04 dataset \cite{wuJointAcneImage2019}. The number of pimples reflects acne severity and we also consider scenarios with different radiuses. Treatment effects are simulated to reduce these acne circles and remove them from a person's image during intervention phases to mimic immediate effectiveness. In each scenario, trials over 56 time points (i.e., images) were simulated for 1000 different individuals, and we evaluated empirical type I error and empirical power of our approach. For the statistical inference, three models were considered, a two-sample t-test of the mean difference of the PC scores, a Wald test of the regression coefficient of treatment in a linear model \ac{AR1}, and a randomization test. 
After an evaluation of our approach in these simulation studies, for illustration, we present an application to the published acne trial \cite{fuMultimodalNof1Trials2023} and compare the findings to the original paper's results of a supervised model. Finally, we perform sensitivity analyses to investigate explainability of our approach. See the Methods section for more details.

\subsection{Simulation Study of Imaging Trials} \label{simstudyimage}

\begin{figure}[ht!]
    \centering
    \includegraphics[width=\textwidth]{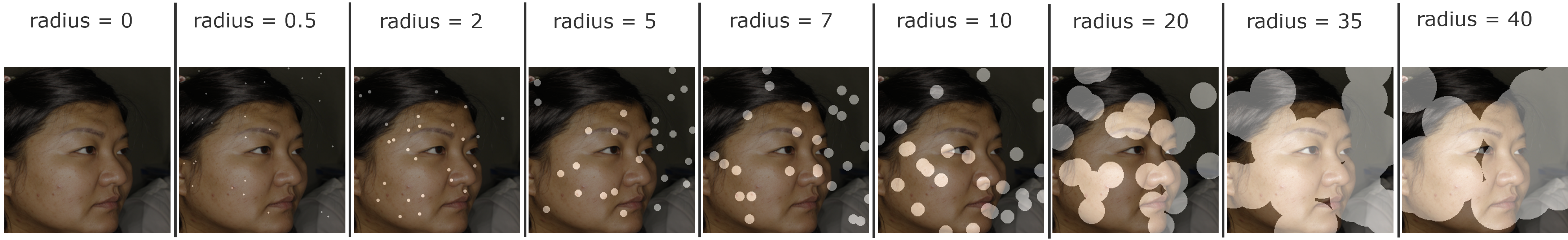}
    \caption{Example images from the simulation study with inserted white circles of different radiuses.}
    \label{fig:simulated_images}
\end{figure}

In the simulation study, we evaluated empirical type I error and empirical power of our approach. Figure \ref{fig:simulated_images} shows an illustration of the generated data underlying the simulation study. Here, we included scenarios with realistic pimple sizes as well as scenarios with unrealistic sizes as sensitivity checks of the autoencoder.

Under the null hypothesis, data was generated so that treatment assignment did not affect the number of generated pimples. The results for different radiuses are shown in Table \ref{table:no_effect_simulation_results} and illustrate that the estimated empirical type I errors were generally very close to the nominal level of 0.05 for all realistic radiuses between 0.5 and 10 and all considered statistical models. 
The error rate remained nearly constant for all tests at varying radiuses.
The error rates for the t-test ranged between 0.038 and 0.050, the error rates for the \ac{LM} ranged between 0.000 and 0.061, and the \ac{RT}'s error rates ranged between 0.038 and 0.050.
Even for larger unrealistic radiuses of 35 and 40 which were rather creating unrealistic artifacts, the estimated type I error was close to the nominal level when the t-test or randomization test was used. For radius 0, the empirical type I error was close to 0. This scenario was chosen also as a proof-of-concept of the behavior of the unsupervised approach, for a situation which would not appear in practical applications, but where we still wanted to ensure that there is no unwarranted behavior. A radius of 0 in our simulation study was reflected in that, unlike for other radiuses, there was no variation at all in the images. Hence, the results show that in this artifact, the unsupervised approach still yields results, but is not able to identify any signals (see also Table \ref{table:strong_effect_simulation_results}), thus still controlling type I error and not yielding any false insights.
In further analyses, we estimated the empirical type I error rate for the same hypothesis tests (t-test, linear model, randomization test) when applied to the acne severity (which was a numerical variable underlying the inserted pimples in the images) directly, without applying an autoencoder and PCA on the images. The results (see Table \ref{table:no_effect_simulation_results}) showed that for these direct interpretable tests, the empirical type I error was also close to the nominal level for all tests. 

\begin{table}[ht!]
\resizebox{\textwidth}{!}{%
\begin{tabular}{cll|lllllllll|}
\cline{4-12}
\multicolumn{1}{l}{} & & & \multicolumn{9}{c|}{\textbf{Radius}} \\ \cline{3-12} 
\multicolumn{1}{l}{} &
  \multicolumn{1}{l|}{{\color[HTML]{43484C} }} &
  {\color[HTML]{43484C} \textit{\begin{tabular}[c]{@{}l@{}}Baseline  \\ (Simulated Severity)\end{tabular}}} &
  {\color[HTML]{43484C} \textit{Radius 0}} &
  {\color[HTML]{43484C} \textit{Radius 0.5}} &
  {\color[HTML]{43484C} \textit{Radius 2}} &
  {\color[HTML]{43484C} \textit{Radius 5}} &
  {\color[HTML]{43484C} \textit{Radius 7}} &
  {\color[HTML]{43484C} \textit{Radius 10}} &
  {\color[HTML]{43484C} \textit{Radius 20}} &
  {\color[HTML]{43484C} \textit{Radius 35}} &
  {\color[HTML]{43484C} \textit{Radius 40}} \\ \hline
\multicolumn{1}{|c|}{{\color[HTML]{43484C} }} &
  \multicolumn{1}{l|}{{\color[HTML]{43484C} \textbf{t-test}}} &
  {\color[HTML]{43484C} 0.058} &
  {\color[HTML]{43484C} 0.008} &
  {\color[HTML]{43484C} 0.049} &
  {\color[HTML]{43484C} 0.050} &
  {\color[HTML]{43484C} 0.048} &
  {\color[HTML]{43484C} 0.048} &
  {\color[HTML]{43484C} 0.047} &
  {\color[HTML]{43484C} 0.038} &
  {\color[HTML]{43484C} 0.040} &
  {\color[HTML]{43484C} 0.046} \\
\multicolumn{1}{|c|}{{\color[HTML]{43484C} }} &
  \multicolumn{1}{l|}{{\color[HTML]{43484C} \textbf{Linear Model}}} &
  {\color[HTML]{43484C} 0.065} &
  {\color[HTML]{43484C} 0.000} &
  {\color[HTML]{43484C} 0.061} &
  {\color[HTML]{43484C} 0.058} &
  {\color[HTML]{43484C} 0.052} &
  {\color[HTML]{43484C} 0.058} &
  {\color[HTML]{43484C} 0.046} &
  {\color[HTML]{43484C} 0.048} &
  {\color[HTML]{43484C} 0.061} &
  {\color[HTML]{43484C} 0.000} \\
\multicolumn{1}{|c|}{\multirow{-3}{*}{{\color[HTML]{43484C} \textbf{Test}}}} &
  \multicolumn{1}{l|}{{\color[HTML]{43484C} \textbf{Randomization Test}}} &
  {\color[HTML]{43484C} 0.049} &
  {\color[HTML]{43484C} 0.000} &
  {\color[HTML]{43484C} 0.046} &
  {\color[HTML]{43484C} 0.046} &
  {\color[HTML]{43484C} 0.046} &
  {\color[HTML]{43484C} 0.047} &
  {\color[HTML]{43484C} 0.050} &
  {\color[HTML]{43484C} 0.038} &
  {\color[HTML]{43484C} 0.041} &
  {\color[HTML]{43484C} 0.047} \\ \hline
\end{tabular}%
}
\caption{Empirical type I error rate of the statistical tests under the null hypothesis, estimated as the rate of rejected tests across the 1000 simulated trials for an $\alpha$ level of 0.05, for each of the varying radiuses.}
\label{table:no_effect_simulation_results}
\end{table}

Next, we estimated empirical power in simulated data where treatment assignment affects the number of inserted pimples in the images.
The results are shown in Table \ref{table:strong_effect_simulation_results} and show that across all radiuses, all tests generally were able to identify a treatment effect with high power
ranging between 0.8 and 1.
Overall, the power did not increase monotonously with the radius of the inserted circles, but was rather constant over radiuses and fluctuated for every test. Again, extreme radiuses used for sensitity checks of 0 and 40 indicated that analyses of high radiuses still yielded high power, while a radius of 0 led to very low power. Analyzing the numerical simulated severity directly with the statistical models yielded similar empirical power compared to our approach using autoencoders and PCA.

\begin{table}[ht!]
\resizebox{\textwidth}{!}{
\begin{tabular}{cll|lllllllll|}
\cline{4-12}
\multicolumn{1}{l}{}                                                         &                                                                          &                                                                                                           & \multicolumn{9}{c|}{\textbf{Radius}}   \\ \cline{3-12} 
\multicolumn{1}{l}{}                                                         & \multicolumn{1}{l|}{{\color[HTML]{43484C} }}                             & {\color[HTML]{43484C} \textit{\begin{tabular}[c]{@{}l@{}}Baseline  \\ (Simulated Severity)\end{tabular}}} & {\color[HTML]{43484C} \textit{Radius 0}} & {\color[HTML]{43484C} \textit{Radius 0.5}} & {\color[HTML]{43484C} \textit{Radius 2}} & {\color[HTML]{43484C} \textit{Radius 5}} & {\color[HTML]{43484C} \textit{Radius 7}} & {\color[HTML]{43484C} \textit{Radius 10}} & {\color[HTML]{43484C} \textit{Radius 20}} & {\color[HTML]{43484C} \textit{Radius 35}} & {\color[HTML]{43484C} \textit{Radius 40}} \\ \hline
\multicolumn{1}{|c|}{{\color[HTML]{43484C} }}                                & \multicolumn{1}{l|}{{\color[HTML]{43484C} \textbf{t-test}}}              & {\color[HTML]{43484C} 1.000}                                                                              & {\color[HTML]{43484C} 0.004}             & {\color[HTML]{43484C} 1.000}               & {\color[HTML]{43484C} 1.000}             & {\color[HTML]{43484C} 0.886}             & {\color[HTML]{43484C} 0.833}             & {\color[HTML]{43484C} 0.881}              & {\color[HTML]{43484C} 1.000}              & {\color[HTML]{43484C} 0.807}              & {\color[HTML]{43484C} 1.000}              \\
\multicolumn{1}{|c|}{{\color[HTML]{43484C} }}                                & \multicolumn{1}{l|}{{\color[HTML]{43484C} \textbf{Linear Model}}} & {\color[HTML]{43484C} 1.000}                                                                              & {\color[HTML]{43484C} 0.000}             & {\color[HTML]{43484C} 1.000}               & {\color[HTML]{43484C} 1.000}             & {\color[HTML]{43484C} 0.883}             & {\color[HTML]{43484C} 0.834}             & {\color[HTML]{43484C} 0.885}              & {\color[HTML]{43484C} 1.000}              & {\color[HTML]{43484C} 0.811}              & {\color[HTML]{43484C} 1.000}              \\
\multicolumn{1}{|c|}{\multirow{-3}{*}{{\color[HTML]{43484C} \textbf{Test}}}} & \multicolumn{1}{l|}{{\color[HTML]{43484C} \textbf{Randomization Test}}}  & {\color[HTML]{43484C} 0.929}                                                                              & {\color[HTML]{43484C} 0.000}             & {\color[HTML]{43484C} 0.945}               & {\color[HTML]{43484C} 1.000}              & {\color[HTML]{43484C} 0.887}             & {\color[HTML]{43484C} 0.837}             & {\color[HTML]{43484C} 0.882}              & {\color[HTML]{43484C} 1.000}               & {\color[HTML]{43484C} 0.817}              & {\color[HTML]{43484C} 1.000}              \\ \hline
\end{tabular}
}
\caption{Empirical power of the statistical tests under the alternative hypothesis, estimated as the rate of rejected tests across the 1000 simulated trials for an $\alpha$ level of 0.05, for each of the varying radiuses.}
\label{table:strong_effect_simulation_results}
\end{table}

In summary, these results suggest that our proposed approach yields valid type I error and is capable of identifying effects of different sizes in imaging trials with high statistical power, for different radiuses and even then artifacts of very large pimples are present.

\subsection{Application in Acne N-of-1 Trials}
\label{applicationacne}

In a next step, we applied our proposed approach to the published series of N-of-1 trials in Fu et al. \cite{fuMultimodalNof1Trials2023}. Using our approach, we were able to reproduce findings using supervised models trained on expert knowledge. Our model successfully identified a treatment effect in participant 2 when the t-test was used, which had shown slightly highest power in the simulation study. 
Table \ref{table:test_results} shows the detailed results and that the t-test identifies a treatment effect in participant 2, which was also shown in the original analyses.
The p-value of the Wald test of the regression coefficient for intervention in the linear model and in the randomization test were both slightly larger than 0.05.

\begin{table}[ht!]
\center
\begin{tabular}{|cl|lllll|}
\hline
\multicolumn{2}{|l|}{\multirow{2}{*}{}}                           & \multicolumn{5}{c|}{\textbf{Participant}}                                                                                                                           \\ \cline{3-7} 
\multicolumn{2}{|l|}{}                                            & \multicolumn{1}{c}{\textit{1}} & \multicolumn{1}{c}{\textit{2}} & \multicolumn{1}{c}{\textit{3}} & \multicolumn{1}{c}{\textit{4}} & \multicolumn{1}{c|}{\textit{5}} \\ \hline
\multicolumn{1}{|c|}{\textbf{}}     & \textbf{t-test}             & 0.551                          & 0.034                          & 0.918                          & 0.612                          & 0.318                           \\ \cline{2-2}
\multicolumn{1}{|c|}{\textbf{Test}} & \textbf{Linear Model}       & 0.700                          & 0.087                          & 0.954                          & 0.495                          & 0.387                           \\ \cline{2-2}
\multicolumn{1}{|c|}{\textbf{}}     & \textbf{Randomization Test} & 0.700                          & 0.056                          & 0.737                          & 0.924                          & 0.448                           \\ \hline
\end{tabular}
\caption{Results of the statistical hypothesis tests of testing the effect of acne cream on acne severity in the imaging acne N-of-1 trial data of Fu et al. \cite{fuMultimodalNof1Trials2023}. Shown are the p-values of the statistical hypothesis tests based on our proposed approach that were performed separately for each participant.}
\label{table:test_results}
\end{table}

\subsection{Sensitivity Analyses}

In the following, we report results from different analyses investigating the unsupervised learning steps in our proposed approach in order to understand and explained the observed true associations in the simulation study and application.

\subsubsection{Explainability of the Learned Embeddings in the Acne N-of-1 Trials} \label{sec2-4-1}

\begin{figure}[h!]
    \centering
    \includegraphics[width=0.8\textwidth]{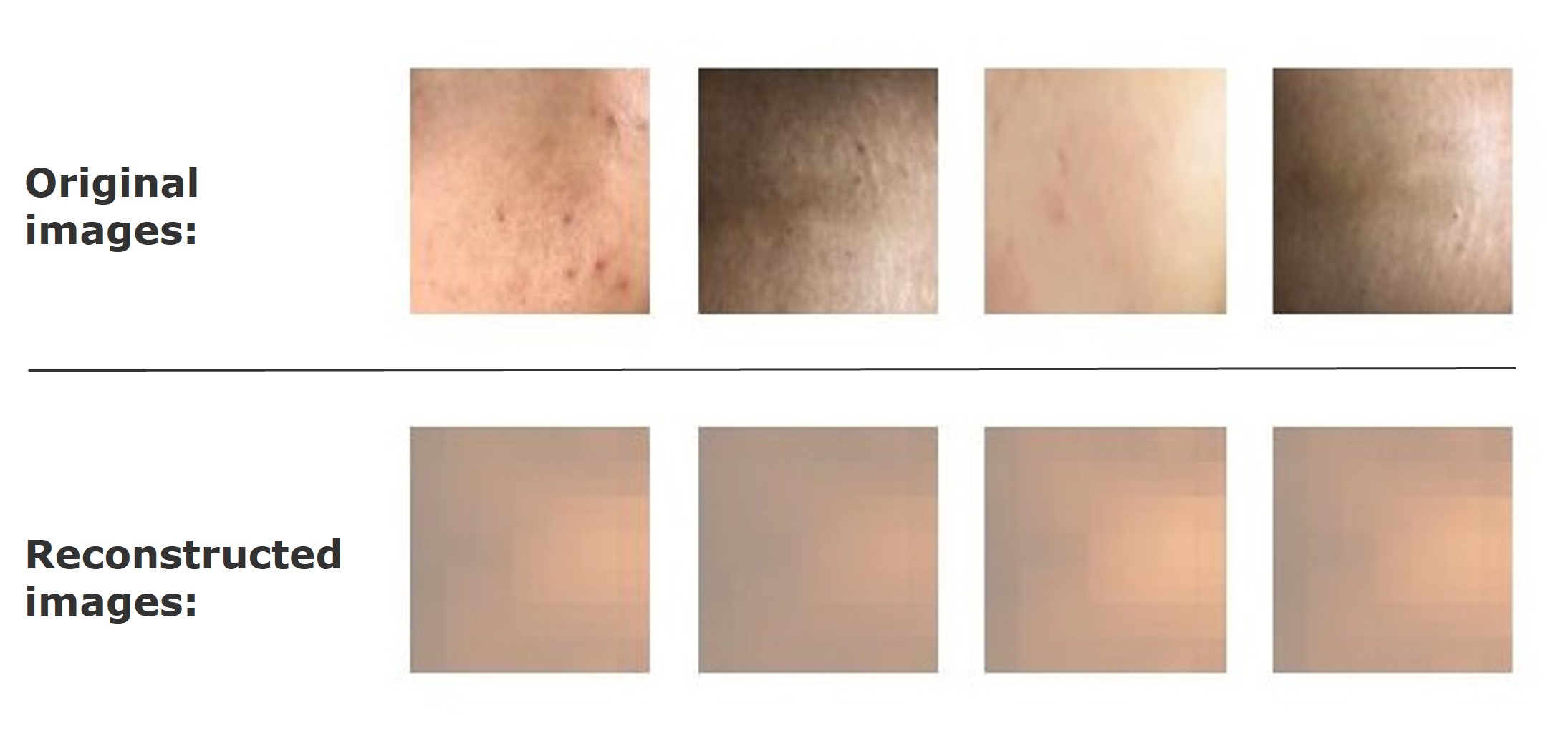}
    \caption{Examples of original images from the acne N-of-1 trials (top panel) and reconstructed images from the autoencoder in our proposed approach (bottom panel).}
    \label{fig:reconstructed_images}
\end{figure}

In the autoenconder of our proposed approach, two models are trained: an encoder capturing the features of the input image in embeddings, and and a decoder that tries to reconstruct the input image based on the embedding vectors. First, we inspected how well the original images can be reconstructed by the decoder. As an illustration, Figure \ref{fig:reconstructed_images} shows a selection of original and reconstructed images in the acne trial. Beige colours and shading are recognizable in the reconstructed images, but they are largely blurry. This is a well-known occurence in autoencoders \cite{kimTwoProposedSolutions2023} and indicates that images cannot be reconstructed well. This does not contradict the validity of the results described in sections \ref{simstudyimage} and \ref{applicationacne}, but highlights the challenges due to the limited data available in the analysis, and hints large potential for improvement in future work. Once embeddings were obtained from the autoencoder, we performed PCA and extracted the first PC. In this step, the first \ac{PC} explained 98.2 \% of variance in the embeddings, which indicates that with the extraction of this single PC (out of the 64-dimensional embedding), almost all of the variation contained in the embedding's latent vectors is explained.

Next, we investigated whether the extracted PC scores (learned from our unsupervised learning approach) mirror the self-rated acne severity scores. Of note, the self-rated acne severity scores were not used in training of the autoencoder and were based on expert ratings (see Methods section for details). For some more background, in the data available from the series of acne N-of-1 trials, each image was manually rated with respect to its acne severity, and the analyses in Fu et al. \cite{fuMultimodalNof1Trials2023} illustrated that there was a treatment effect in participant 2 with respect to this acne severity score. One question is whether the PC extracted from our proposed approach closely mirrors this acne severity score (since we also observed a treatment effect in participant 2), or whether the observed effect was rather based on other characteristics that were extracted from the images in the autoencoder.  

\begin{figure}[H]
\includegraphics[width=.8\textwidth]{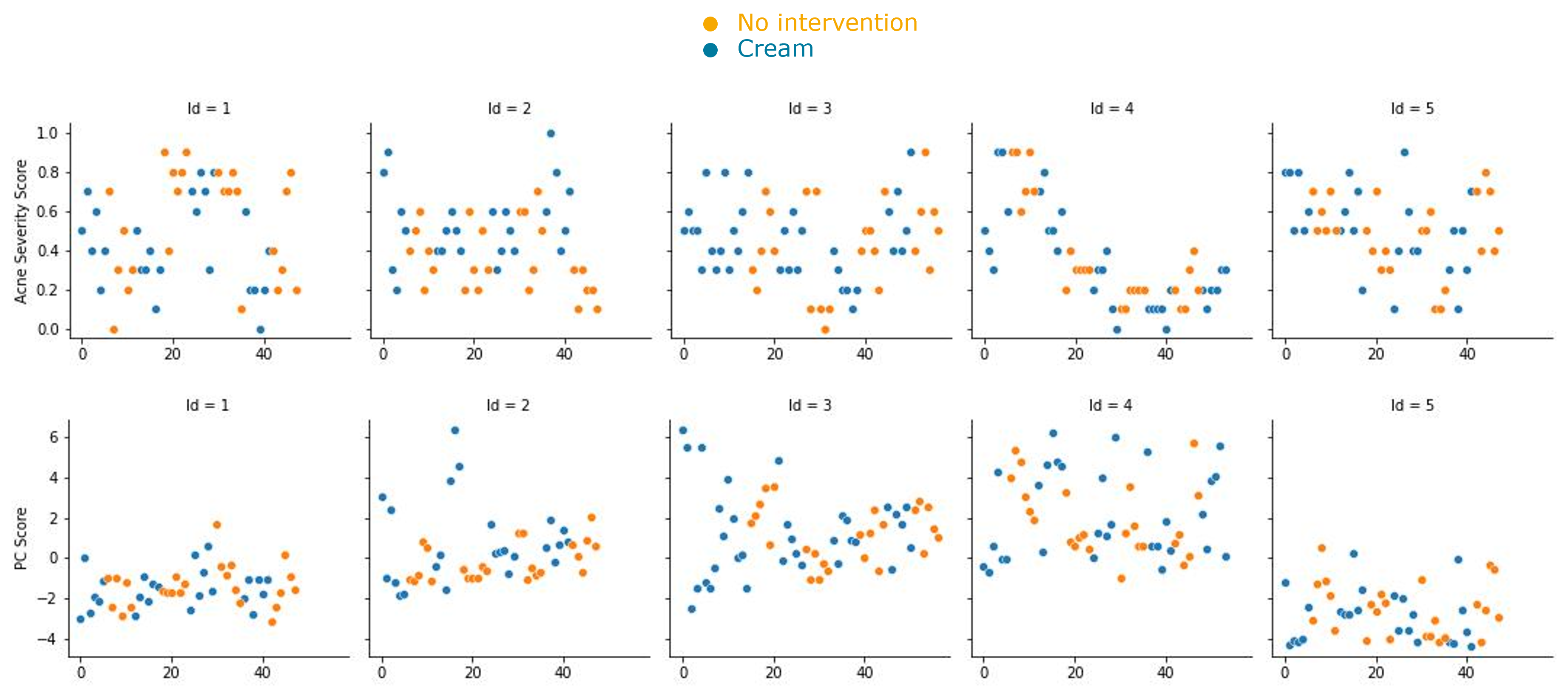} 
\centering
\caption{Time series plots of the acne severity scores (top panel) and \ac{PC} scores (bottom panel) over time, for each participant ("ID") separately, in the acne N-of-1 trial. Non-intervention phases are highlighted in orange, intervention phases are highlighted in blue.}
\label{fig:all_timeseries}
\end{figure}

Figure \ref{fig:all_timeseries} shows the time series for the manual acne severity scores and the \ac{PC} scores for each ID. While the average differences between cream and no cream phases are visible for participant 2, they are not fully aligned, which is an indication that the embeddings from our unsupervised approach pick up other or additional features in the images. For the other participants, some similarities are visible but no general trends.
Overall, the estimated correlations between the acne severity scores and the PC scores obtained from our approach were small (Person correlation r=0.30 for participant 1, r=0.14 for participant 2, r=0.01 for participant 3, r=0.24 for participant4, and r=0.02 for participant 5). The small association is also illustrated in Table \ref{table:ranges_avgscore_pca2}, which shows the average acne severity scores and average PC scores across conditions and participants.

\begin{table}[H]
\centering
\resizebox{\textwidth}{!}{%
\begin{tabular}{cl|lllll|}
\cline{3-7}
\textbf{}                                                     & \multicolumn{1}{c|}{\textbf{}}             & \multicolumn{5}{c|}{\textbf{Participant}}                                                                                                                           \\ \hline
\multicolumn{1}{|c|}{\textbf{Score}}                          & \multicolumn{1}{c|}{\textbf{Intervention}} & \multicolumn{1}{c}{\textit{1}} & \multicolumn{1}{c}{\textit{2}} & \multicolumn{1}{c}{\textit{3}} & \multicolumn{1}{c}{\textit{4}} & \multicolumn{1}{c|}{\textit{5}} \\ \hline
\multicolumn{1}{|c|}{\multirow{2}{*}{\textbf{Acne severity}}} & \textbf{Cream}                             & 0.54 (0.28)                    & 0.36 (0.18)                    & 0.43 (0.23)                    & 0.37 (0.27)                    & 0.48 (0.19)                     \\
\multicolumn{1}{|c|}{}                                        & \textbf{No cream}                          & 0.42 (0.23)                    & 0.54 (0.20)                    & 0.46 (0.20)                    & 0.34 (0.25)                    & 0.52 (0.22)                     \\ \cline{1-2}
\multicolumn{1}{|c|}{\multirow{2}{*}{\textbf{PC Score}}}      & \textbf{Cream}                             & -1.32 (1.04)                   & -0.15 (0.94)                  & 1.20 (1.40)                    & 1.91 (1.81)                    & -2.54 (1.36)                    \\
\multicolumn{1}{|c|}{}                                        & \textbf{No cream}                          & -1.50 (0.96)                   & 0.86 (2.05)                    & 1.25 (2.17)                    & 2.20 (2.25)                    & -2.93 (1.32)                    \\ \hline
\end{tabular}}
\caption{Means and standard deviations of the acne severity scores and the \ac{PC} scores in the acne N-of-1 trial, stratified by participants and intervention.}
\label{table:ranges_avgscore_pca2}
\end{table}

\subsubsection{Explainability of the Learned Embeddings in the Simulation Study}

\begin{figure}[h!]
    \centering
    \includegraphics[width=0.8\linewidth]{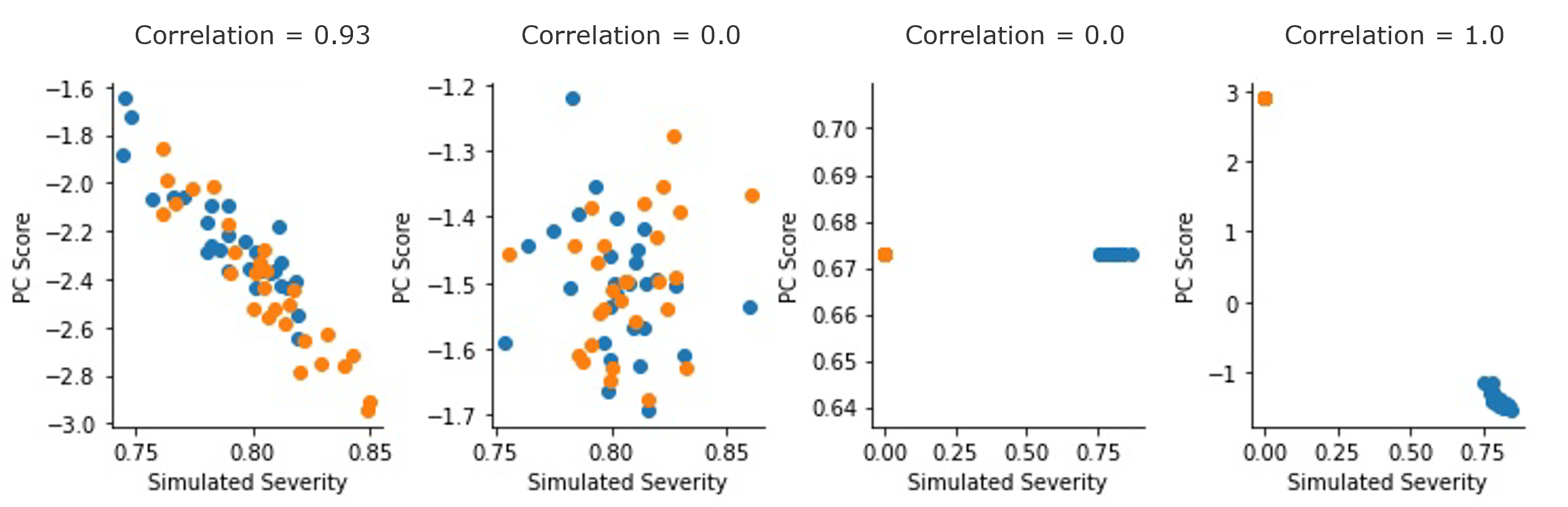}
    \caption{Scatterplots of acne severity (x-axis) and PC scores from the autoencoder (y-axis) for different scenarios and different participants in the simulation study. Shown are the scatterplots for the situations where highest and lowest correlations were observed between acne severity and PC scores, respectively. The two leftmost panels show correlations in the scenario without effect (under the null hypothesis). Likewise for the two rightmost panels for the scenario with effect (under the alternative hypothesis).}
    \label{fig:ind_corrs}
\end{figure}

After inspecting which features of the images have been learned by the autoencoder in the real imaging N-of-1 trials, we performed further sensitivity analyses in the simulation study.
First, we inspected the model quality in the different scnarios, which indicated that the loss in the autoencoder was similar, and consistently small across all scenarios. Also, the first PC explained a large amount of variance of the information captured by the embedding's latent vectors. The explained variance was higher than 70\% in most scenarios and ranged between 50\% and 99.8\%.
Next, we inspected the association of the PC scores with the acne severity scores (which were used in the generation of the images in the simulation study) to investigate which features were learned by the autoencoders in the simulation study. The results indicate that the embeddings learned in the autoencoder are highly correlated with the acne severity scores of the images in some scenarios, but go beyond the acne severity scores of the images and capture more characteristics of the images in some other scenarios. This is illustrated in Figure \ref{fig:ind_corrs}, which shows scatterplots of the acne severity scores and PC scores for some exemplary scenarios and some participants. When the association between acne severity scores and PC scores was investigated across participants, the Pearson correlation between the acne severity scores and PC scores was below r=0.1, indicating that careful interpretation needs to be made on the individual level.

\section{Discussion} \label{Discussion}

In this work, we present an approach that combines unsupervised learning with statistical inference for an automated analysis of multimodal outcome data in N-of-1 trials. In the unsupervised learning part, we train an autoencoder to create efficient embeddings of input images of N-of-1 trials that are subsequently reduced to one dimension through principal component analysis, which is then tested for its association with treatment. 

After presenting the approach, we first evaluated it in imaging simulation studies for which we simulated treatment effects onto acne severity by inserting pimples on images of faces of the ACNE04 dataset \cite{wuJointAcneImage2019} as a heuristic for acne symptoms. We generated scenarios without treatment effect in which treatment did neither affect the baseline simulated severity nor the mean number of pimples, and we also generated scenarios with treatment effect in which both the simulated severity and the number of inserted pimples were set to zero during treatment phases. The scenarios included nine different sizes of pimples and also checked for extreme values to uncover the behavior of our proposed approach in edge cases and possible artifacts. To demonstrate feasibility and ease interpretation of the results, we did not consider time dependencies and carryover effects of treatment.
In the analysis, the loss in the autoencoder training was consistently small across all simulated scenarios. The results of the statistical tests based on the \ac{PC} scores from the simulated images' embeddings showed that our approach allows inference on treatment effects with valid empirical type I error rates and high statistical power. Of note, the statistical power to identify treatment effects was high when treatment effects was large in the form of large pimples, but also for medium-sized and very small pimple differences between treatment and no treatment.
We observed that the estimated statistical power was varying in a non-monotonous way across scenarios with different pimple radiuses, which may be attributed to the model becoming more sensitive to background changes in the images for smaller pimples. When pimples with a radius of 0 were inserted, i.e., instead of white circles, a noise was introduced at random pixels in a given image, the minuscule differences that thus existed between images of treatment and non-treatment phases yielded empirical type I errors of 0.8\% when there were no treatment effects and and empirical power of 0.4\% when there was a treatment effect using the t-test, and 0\% identified treatment effects across all scenarios for the linear model and randomization test. That is, while these single-pixel changes in images do not capture enough information to allow an identification of treatment effects, they also do not lead to erroneous conclusions and inflated type I errors.
In a second analysis, we applied our approach to data from a real-world empirical multimodal N-of-1 trial. Our approach identified a treatment effect for one of the participants, thereby yielding the same conclusions as the direct analysis of self-reported acne severity scores and as an analysis based on supervised deep learning models reported in \cite{fuMultimodalNof1Trials2023}. This indicates that our unsupervised approach is successful in replicating the results of supervised approaches, which require manual expert labeling with supervised models and are not applicable on a wider scale of N-of-1 trials.

Following the demonstration that our proposed approach satisfies desired statistical guarantees and is able to identify treatment effects in real-world data, we performed several follow-up analyses to uncover which information of the images is captured by the unsupervised learning. This can be important for follow-up analyses and interpretations in order to understand which skin differences might be underlying any identified treatment effects. A first question is how closely the extracted embeddings reduced to a single dimension using principal component analysis mirror numeric measures of acne severity, or how much additional information they capture from the images. 
In the context of the simulation study, the explained variance of of the information captured by the embedding's latent vectors by the PC scores varied between 0.4 and 1 across the different scenarios. The correlations between acne severity and \ac{PC} scores was small for scenarios without treatment effects and varied extensively from very small to very large in scenarios with treatment effects. This could be interpreted as variability in captured features between the scenarios. Moreover, differences in the means of the \ac{PC} scores between scenarios with and without treatment effects indicate that our proposed approach is able to capture differences in the outcome images when there is a treatment effect and when there is no effect.
In the context of the application to the real-world acne N-of-1 trial data, the reconstructed images were blurry and showed shades of skin colors. This is a common occurrence in generative models, which often lead to blurry reconstructed images. Although the reconstructions were not of considerable quality, it can still be assumed that the embeddings captured meaningful information about the images, as demonstrated by the findings matching those of the original acne pilot study.
While the correlation between acne severity scores and \ac{PC} scores was small or moderate, the \ac{PC} score (i.e., first principal component) explained 98.2\% variance of the embeddings, so it captures the features of the encoder model very closely.

In a comparison of the applied statistical tests, all tests yielded generally similar results, but only the t-test yielded a p-values smaller than 0.05 in participant 2 in the acne application, while the linear model and randomization test yielded p-values higher than 0.05. Sensitivity analyses indicated that the residuals did not follow an AR1 structure but only showed a very small dependence, which in turn might have yielded the small observed efficiency loss in the misspecified model of the linear model compared to the t-test.

Following the results presented here, we see several avenues for improvement and future work.\\
First, for training of the autoencoder, we used standard hyperparameters and standard image augmentation techniques. More sophisticated hyperparameter or architecture tuning may improve the reconstruction results. Other models such as variational autoencoders might also yield more meaningful embeddings. Similarly, more elaborate statistical models and hypothesis tests could incorporate time-dependencies, covariates, and carryover effects to test more complex treatment effects \cite{liaoAnalysisNof1Trials2021, vieiraDynamicModellingNof12017, dazaModelTwinRandomizationMoTR2022}.\\
Second, when training deep learning models, the amount of training data available is crucial for training a performant model, in our case to extract meaningful features in the embeddings of the autoencoder. Depending on the specific N-of-1 trial, suitable openly available datasets for pretraining or proxy-task training of a model to later be fine-tuned on the trial data may be helpful.\\
Third, not only the amount, but also the quality of data is essential. The acne severity scores in the application of the data from \cite{fuMultimodalNof1Trials2023} could be improved by collecting more ratings from a larger group of experts (or non-experts) to reduce variance and increase inter-rater-reliability. Furthermore, applying acne scoring methods based on severity and number of lesions on the skin, as have been developed already and are used by dermatologists, may help generate reliable labeled data or be used as a training task for the model.\\
Fourth, in the autoencoder architecture, we created a 64-dimensional embedding layer and then extracted one dimension from the latent embedding using \ac{PCA}. As an alternative, the outcome dimensionality could be reduced by pushing the autoencoder to compress the input to one-dimensional or lower-dimensional embeddings directly. Other orders of dimensionality of the latent space may yield better results, or other, possibly non-linear, dimension reduction methods such as t-SNE \cite{maatenVisualizingDataUsing2008} could be employed. As further option, the \ac{PCA} does not need to be restricted to only one factor, and could be skipped entirely by performing multivariate hypothesis tests such as proposed in \cite{article:kirchler20a}.\\
Lastly, further investigations into the clinical interpretation of the embeddings given the low amount of data in a single N-of-1 trial will be required. At present, the question remains as to how the meaningfulness of the latent space can be assessed optimally. We suggest different types of visualizations and comparisons with trusted baseline results or simulation studies.\\
Whereas our proposed unsupervised learning approach yielded results that agreed with our expectations both in simulated and for real-world data, this does not seem to be attributable to the produced \ac{PC} scores recreating the underlying acne severity, as we initially assumed. This is evident in the variation in correlation between \ac{PC} scores and acne severity scores. Large differences in explained variance between the \ac{PC} scores and the means of the \ac{PC} scores point towards complex relationships that need follow-up investigation. In particular for future automated clinical applications, explainability of the embeddings will be crucial.

In summary, we believe that our proposed approach can serve as a strong baseline and provide a valuable step towards incorporating unsupervised models on multimodal outcome data in N-of-1 trials. This will be crucial to enable the integration of multimedia data into N-of-1 trials towards more accessible and comprehensive personalized interventions that may serve a broader population in the future.

\section{Methods} 
\label{Methods}

The overall aim of this work is to provide a framework for automatically identifying effects of one or more treatments on an outcome, assessed by unstructured data such as images, where no labels are provided.\\
In order to use images directly as outcome data, models need to be trained that can find meaningful embeddings capturing the relevant features in the image, in our case visual properties of acne on the skin.
Unsupervised learning approaches can be used in this setting to compress relevant information of an image in a lower-dimensional space.
There is a variety of available unsupervised approaches, which differ not only in their architecture but also in their training methods. For instance, pretrained models could be fine-tuned for a given specific task, or a proxy-learning task may enhance the power of a model.\\

Here, we use a three-step approach to represent an image in a 1-dimensional variable that is used in follow-up statistical hypothesis testing. First, we use an autoencoder to embed the images in a lower-dimensional embedding space to capture the relevant features of acne severity in an unsupervised fashion. Autoencoders can represent complex information in the embedding space and can be trained without labels \cite{baldiAutoencodersUnsupervisedLearning}. Second, we use \acf{PCA} to reduce the dimensionality of the embeddings to one dimension, which can be used for statistical testing. In the third step, we apply different univariate hypothesis tests, namely a t-test, a Wald test of the regression coefficient in a linear model \acf{AR1} and a \acf{RT} to test the null hypothesis of no difference in outcomes between intervention and non-intervention phases. See Figure \ref{fig:approach} for an illustration.

\subsection{Approach}

\subsubsection{Step 1: Unsupervised Learning to Obtain Embeddings of Outcomes}

In step 1, we use autoencoders as an efficient tool to learn lower-dimensional representations of an input. Autoencoders consist of two deep models: an encoder capturing the features of an input -- here, an image -- and a decoder that is provided the latent space vectors from the encoder to reconstruct the input \cite{Goodfellow-et-al-2016}.
By transforming the original input, the autoencoder maps into a lower-dimensional space and can therefore create efficient nonlinear, possibly correlated representations. The encoder poses a bottleneck for the decoder, as it aims to represent the input efficiently and capture relevant features, while not creating exact copies \cite{Goodfellow-et-al-2016}.
Since typical N-of-1 trials supply relatively few data points, we hypothesize that this bottleneck property may help with accurately embedding the input features. The restored images provide a visual heuristic to evaluate the quality of the embeddings.

Elaborating further, autoencoders comprise two functions, $T = G(Y)$ and $Y' = H(T)$, where $Y$ is the input image, $G(\cdot)$ is an arbitrarily complex mapping function into the latent space, $T$ is the latent space embedding, and $H(\cdot)$ is a mapping function to the restored input $Y'$, with the obective to minimize the distance between $Y$ and $Y'$
\cite{kramerNonlinearPrincipalComponent1991}. Thus, $G(\cdot)$ represents the encoder part and $H(\cdot)$ represents the decoder part of the model.

In this study, we specified the loss for the two-stage optimization problem as the sum of the reconstruction errors, i.e. the mean squared error between reconstructed and original image \cite{kramerNonlinearPrincipalComponent1991}. 
The loss $L$ is thus:
$L = \frac{1}{n}\frac{1}{w}\sum_{i=1}^{n}\sum_{p=1}^{w}(Y_{ip} - Y'_{ip})^{2}$,
where $Y_{ip}$ denotes the pixels $p=1,...,w$ of the respective images $i=1,...,n$ and their reconstructions. In our analyses, we use the images of all participants in training, that is, $n$ is the product of participants and number of images (time points) per participant. 
As a result of training the autoencoder, we extract the $k$-dimensional latent embedding, $k\leq n$, where we set $k$ to 64.

\subsubsection{Step 2: Principal Component Analysis to Reduce the Dimensionality of Embeddings}

In step 2, we use \acf{PCA} as a popular technique for dimensionality reduction. PCA performs an orthogonal linear transformation of high-dimensional input data that can be used to project into a lower-dimensional space while maintaining as much variation of the original data as possible.
Let $n$ be the number of images and $k$ the dimensionality of the latent variables $X_i$ in the embeddings in the autoencoder from step 1, and let $\textbf{X} =(X_1, ..., X_k) \in \mathbb{R}^{n\times k}$. Then with \ac{PCA}, we linearly transform the $k$ latent variables into a single new variable $Z = \textbf{X}\omega$, $\omega \in \mathbb{R}^k$, with the aim of maximizing $var(Z)$ over $||\omega||=1$. 
Here, $||\cdot||$ is the Frobenius norm, since arbitrarily large $\omega$ would result in an arbitrarily large $var(Z)$ \cite{broPrincipalComponentAnalysis2014}.
\acf{PC}s are the eigenvectors of the input's covariance matrix, and as such can be computed by eigenvalue decomposition or singular value decomposition \cite{jolliffePrincipalComponentAnalysis2016}. Thus, $var(Z)=var(\textbf{X}\omega) = \omega'\textbf{S}\omega$, where \textbf{S} is the variance-covariance-matrix of $\textbf{X}$. The maximization problem then can be rewritten as:
%
   $\arg \max_{||\omega||=1} \omega'\textbf{S}\omega - \lambda(\omega'\omega -1)$,
%
where $\lambda$ is the Lagrange multiplier. Differentiating and solving with respect to $\omega$ yields that $\omega$ is the eigenvector and $\lambda$ the corresponding eigenvalue of \textbf{S}.
In our approach, the latent vectors of the embeddings from the autoencoder serve as input data for the \ac{PCA}. Accepting a loss of information, we extract only the first \ac{PC} to enable direct use of classical statistical tests in estimating the empirical power and type I error in simulation studies and for a direct comparison to the analysis of the acne severity scores provided in the study of Fu et al. \cite{fuMultimodalNof1Trials2023}.  

\subsubsection{Step 3: Statistical Inference on Treatment Effects with PC Scores}\label{Hypothesis Testing}

In step 3, we test whether the embeddings, generated from the input images, are associated with treatment using classical statistical tests. We use classical statistical tests -- t-tests, linear models and randomization tests -- that provide a straightforward interpretation. Interpretability is especially useful when involving clinicians or patients in treatment decisions based on the trial outcomes. 

As first hypothesis test, we are using a two-sample t-test, which compares the expected values of two groups from the same population. 
Let $I$ be a binary variable for intervention which is 1 under treatment and 0 under no treatment, and
let $PC$ be the the \ac{PC} score resulting from the autoencoder's embeddings. Then the null hypothesis for the t-test in the acne trial is $H_{0}: E(PC|I = 1) = E(PC|I = 0)$ and the test statistic is
%
			$t = \frac{\sum_{i=1}^{n} D_{i}}{\sqrt{s^{2} (\frac{1}{n_{I=1}} + \frac{1}{n_{I=0}})}}$
%
with $n_{I=1}$ and $n_{I=0}$ as the respective sample sizes of measurements in intervention or non-intervention phases and
    $s^{2} = \frac{\sum_{i=1}^{n_{I=1}}(PC_{i}-\overline{PC}_{I=1})^{2} + \sum_{j=1}^{n_{I=0}}(PC_{j}-\overline{PC}_{I=0})^{2}}{n_{I=1} + n_{I=0} -2}$
%
as pooled variance, where we set $ D_{i} = \overline{PC}_{I=1}-\overline{PC}_{I=0}$ as the difference in means between intervention and non-intervention phases. Here, $i$ is the $i$th observation of $n$ total observations in either intervention or non-intervention phases. We assume in this case that the variance $s^{2}$ is equal for both samples.

For the second hypothesis test, which we will call abbreviated "linear model", we are assuming a linear regression model, which has been another popular choice for the analysis of N-of-1 trials \cite{kravitzDesignImplementationNof1, shafferNof1RandomizedIntervention2018} and allows to incorporate the influence of covariates. The general model equation is
%
	$PC_{t} =\beta_{0} + \beta_{1}I_t + \epsilon_{t}$,
%
where $PC_{t}$ is the \ac{PC} at time point $t$, and we are testing the treatment effect by evaluating the null hypothesis $H_{0}:\beta_{1} = 0$ using Wald tests. We assume that the error term $\epsilon_{t}$ follows a first order autoregressive (AR1) structure with $\epsilon_{t} = \rho\epsilon_{t-1} + \varepsilon_{t}$ and $\varepsilon_{t} \sim \mathcal{N}(0, \sigma^2)$.
Incorporating an AR1 error structure acknowledges possible time dependencies of the measurements in a single-person trial \cite{StateArtFuture}. The structure is such that the covariance between the errors decreases towards zero with increasing lags.

As a third hypothesis test, we applied permutation tests as a non-parametric approach. An advantage of permutation tests is that they do not rely on assumptions of normality compared to the t-test and linear model. In permutation tests for two treatment groups -- i.e., intervention and non-intervention -- the realized partition of outcome values to the two groups are compared to all possible partitions and then the p-value of the realized outcome is calculated. 
A bit more formally, let us consider our data with sample size $N$ (i.e., $N$ images of one person across all time points in the study), that can be randomly split into two groups of sizes $q$ and $r$, which yields $\binom{N}{q}$ possible assignments to the two groups \cite{ernstPermutationMethodsBasis2004}.
We test the hypothesis that the outcome data from intervention and non-intervention phases within one person stem from the same distribution \cite{heyvaertRandomizationTestsSinglecase2014}, that is, we test the null hypothesis \cite{bouwmeesterPowerRandomizationTest2020} that $H_{0}: E(PC_j|I = 1) = E(PC_j|I = 0)$ for all permutations $j = 1, 2, ..., {\binom{N}{q}}$, where $PC_j$ denotes the PCs in the $j$-th permutation. 
Still, it is necessary to take into account the order of the phases within a person's trial, since the data are not independent. Thus, in a \ac{SCRT}, the realized ordered outcome of the trial at hand (i.e. the order of treatment and non-treatment phases and their respective outcome values) is compared to all possible orders of assignments of treatment and non-treatment in a trial of the same length \cite{onghenaRandomizationTestsPermutation2017}.
Afterwards, the test statistic of interest for the realized trial is computed. In the case of our study, $E(PC|I=1) < E(PC|I=0)$ is estimated, as the acne cream is supposed to lower the score on an acne severity scale. The realized statistic is now compared to the distribution of all possible assignments of the values and their respective p-value, $p$, is computed.
Now, the p-value for the \ac{RT} is calculated as 
			$p = P(S \le s^{*}|H_{0}) = \frac{\sum_{i=j}^{{\binom{N}{q}}}\mathbf{1}_{(s_{j} \le s^{*})}}{{\binom{N}{q}}}$,
%
where $s_{j} =  \overline{PC}_{j,I=1}-\overline{PC}_{j,I=0}$ is the value of the $j$th test statistic of $m$ randomizations, $\mathbf{1}$ is the indicator function, and $s^{*}$ is the observed value for $s$.
Since computing all possible $\binom{N}{q}$ permutations may be infeasible with larger $N$, Monte-Carlo sampling from the randomization distribution poses a slightly less efficient, but equally exact estimate of the p-value, $\hat{p}$. With $M$ samples, the estimate would then be, following Ernst \cite{ernstPermutationMethodsBasis2004}:
%
	$\hat{p} = \frac{1 + \sum_{i=1}^{M}\mathbf{1}_{(s_{j} \le s^{*})}}{M+1}$.

\subsection{Simulation Study} \label{Methods - Simulation Study}

In order to evaluate the statistical properties of our proposed approach under controlled conditions, we performed a Monte Carlo simulation study. In this study, we generated images with varying degrees of acne severity conditional on the presence of a treatment, both under scenarios of treatment effects and under no treatment effects, and then used those images in order to estimate the empirical power and type I error. More details are provided in the following.

\subsubsection{Data Generation and Simulation Setup} 

In the simulation study, we generated data under the null hypothesis of no treatment effect, and under the alternative hypothesis of a treatment effect. In each of these two cases, we considered different scenarios of nine different acne radiuses. And for each of the 18 possible combinations of null/alternative hypothesis and acne radius, we generated data for 1000 trials (i.e. data for a series of N-of-1 trials with 1000 participants), where each trial had 56 time points (which could be consecutive days of measurements if one image was taken each day) in an ABAB design with each period having 14 time points. The number of 56 time points (images) per trial was chosen to investigate specifically the statistical properties in trials such as the Acne trial application, which are a realistic number of observations in practical applications in general.
As a basis, we first selected images of 1000 individuals from the ACNE04 dataset \cite{wuJointAcneImage2019}. It contains 1,457 images of individuals with acne ranging from severity levels 0 through 3. The severity scores are based on experts' estimations. We randomly selected images of 1000 persons with acne level 0 or 1. The original sizes of the images varied between (194, 259) and (5184, 3456). They were therefore all resized to (256,256) before further alterations were made. 
Next, for each simulated trial in each scenario, we inserted white circles into random places of the image. The number of circles was representing the acne severity of a prespecified level 
and we simulated the data adapting the algorithm from \cite{gartnerComparisonBayesianNetworks2023}. We provide more details in the following.

\begin{figure}[ht!]
    \centering
    \includegraphics[width=0.8\textwidth]{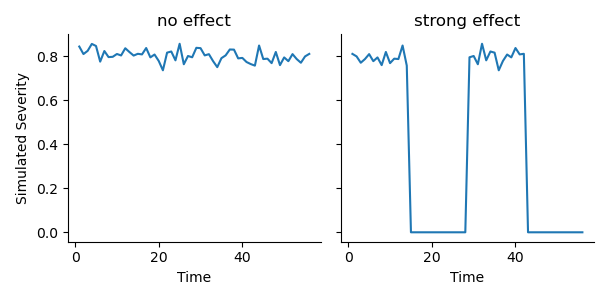}
    \caption{Illustration of the simulated acne severity over time under the null hypothesis and under the alternative hypothesis. This would be the underlying acne severity used in selecting the number of pimples that are further used to drawn onto the face images.}
    \label{fig:simulated_severity}
\end{figure}

For the simulated acne severity, in all scenarios, we chose a constant mean baseline severity level of 0.8 for each of the 1000 participants. 
For the exact number of inserted white circles to eventually vary slightly between images (as is the case with the number of pimples in actual acne), we added a random noise of $\mathcal{N}(0,\,0.025)$ to the severity level for each image, such that the severity level slightly varied around 0.8 for all images. Under the alternative scenarios when there is a treatment effect, we set the simulated severity to 0 during intervention phases, without adding any noise. Formally, this is
\begin{equation}
    o_{t, I} = b + \epsilon_{t} 
\end{equation}
under the null hypothesis, for all time points $t=0,1,...,T$, where \textit{o} is the acne severity, $I$ is the indicator variable for intervention, $b=0.8$ is the baseline severity level and $\epsilon \sim \mathcal{N}(0,\,0.025)$. Note that the equation is the same for both non-intervention and intervention phases under the null hypothesis.
In all scenarios under the alternative hypothesis, the equations are not equal for non-intervention ($I=0$) and intervention ($I=1$) phases:
\begin{equation}
            o_{t,I=0} = b + \epsilon_{t} \quad \text{and} \quad 
            o_{t,I=1} = 0
\end{equation}
This is visualized in Figure \ref{fig:simulated_severity}, where the the simulated acne severities under the null and alternative are shown, respectively.

After defining the underlying acne severity, for each image (for every participant in every scenario), the number of inserted white circles per image was calculated dependent on the acne severity level.
We counted the number of distinct severity levels of each person due to noise, then divided the number by 4 to separate the values into $\frac{\text{number of severity levels}}{4}$ numbered bins, where the lowest numbered bin was assigned the number 20 and the highest numbered bin was assigned the number 33 for every individual. Therefore, the number of white circles inserted into an individual's image varied between 20 and 33 circles with underlying severity level of $0.8 + \mathcal{N}(0,\,0.025)$ under null hypothesis and under the alternative hypothesis during non-intervention phases, whereas under the alternative hypothesis during intervention phases it was 0.

Next, in order to insert the white circles onto the images, a black mask with the size of the original image was created, and white circles were inserted into the mask. The original image and the mask were then blended with a ratio of 60:40 before model training.
The placement of the white circles on the image was randomized. The radius of the white circles was first set to 40 (pixels) to showcase a clear difference in intervention and non-intervention phases. Then, to demonstrate that the autoencoder model focuses on learning the difference in the number of circles, i.e. number of pimples, we gradually reduced the radiuses to 35, 20, 10, 7, 5, 2, and 0.5. We assumed that our approach detects effect differences even in small white circles sizes since in real-world scenarios, pimples may have vanishingly small radiuses compared to total image size and changes in the skin due to treatment might be subtle yet effective.
The radius of 0 was included as a sensitivity check. Due to the way the circles are inserted, a radius of 0 should only change a maximum of one pixel, therefore resulting in minimal noise to the images. We thus would expect similar estimates for the power and for the type I error rate in this case.

\subsubsection{Statistical Analysis} \label{Methods - 1 Statistics}

\begin{figure}[!ht]
    \centering
    \includegraphics[width=0.6\textwidth]{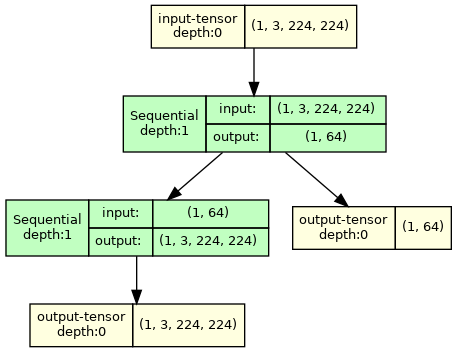}
    \caption{Model architecture of the autoencoder.}
    \label{fig:torchviz}
\end{figure}

The pipeline was constructed with the following steps: First, the input images were preprocessed and augmented using standard approaches: horizontal flips, increase of colour brightness, and a resizing to (224,224) for the acne pilot study as in the original paper and (256,256) for the acne simulation study.
Next, an autoencoder was trained with a batch size of 32 that consisted of 7 convolutional layers and a linear output layer each with ReLU activation in the encoder, yielding the final 64-dimensional output embeddings. The decoder consisted of 1 linear and 5 pairs of convolutional layers and transposed convolutional operators and ReLU activation with a sigmoid function in the end. See a schematic depiction of the autoencoder in Figure \ref{fig:torchviz}. The autoencoder was then trained for 10 epochs with a learning rate of 0.001, using Adam as optimizer. We specified the loss as the reconstruction error, i.e. the mean squared error between reconstructed and original image. Due to the limited amount of data, we forwent a train-test-split, instead using an augmented version of the input images as train and a non-augmented version as validation set.

Afterwards, the 64-dimensional embedding vectors were run through a \ac{PCA}, and the first \ac{PC} was extracted to reduce the data. The resulting \ac{PC} scores were used in the statistical hypothesis tests. All steps were performed in \texttt{python}. In particular, all embeddings including data augmentation were created using \texttt{pytorch} \cite{pytorch}. 
For the \ac{PCA}, \texttt{sklearn} \cite{scikit-learn} was used. 

Subsequently, we applied the different statistical tests described in section \ref{Hypothesis Testing} on the \ac{PC} scores as well as for the acne severity scores for each individual, respectively. The t-tests were computed using \texttt{scipy.stats} \cite{2020SciPy-NMeth} in \texttt{python}. The linear model was fitted using the \texttt{nlme} \cite{nlme} library in \texttt{R} with default settings. 
In addition, we evaluated a permutation test available in the \texttt{scipy} \cite{2020SciPy-NMeth} package in \texttt{python} that is not specific to single case designs. This implementation tests whether the values from the intervention and non-intervention phases were drawn from the same distribution, without taking into account the ABAB structure of the data. Instead, the phases are treated as in a regular crossover trial. Here, we tested the null hypothesis that the expected value of the outcome under treatment \textit{A} is larger than under treatment \textit{B}, $\mu_A - \mu_B$, i.e. $\mu_{O_{I=0}} - \mu_{O_{I=1}} > 0$, or $\mu_{O_{I=0}} > \mu_{O_{I=1}}$, where $\mu_{O_{I=1}}$ describes the expected value of the outcome when the patient was exposed to the treatment and $\mu_{O_{I=0}}$ when the patient was not exposed respectively. 

Finally, we calculated the percentage of p-values smaller than 0.05 in order to estimate the empirical type I error rate (in the scenarios under the null hypothesis) and the empirical power (in the scenarios under the alternative hypothesis), for each scenario and each hypothesis test.

\subsection{Application in Acne N-of-1 Trials} \label{Methods -  Application}

We applied our approach to the acne pilot study data to compare our unsupervised method to the manual labels provided and the corresponding supervised approaches. We describe the data collection process for the acne pilot study in the following.

\subsubsection{Data} \label{Data}

After evaluating the statistical validity and characteristics of our proposed approach in controlled conditions within the simulation study, we applied it onto real-world data to evaluate its performance there. For this aim, we used the data of the published multimodal pilot series of N-of-1 trials on acne (\url{https://github.com/HIAlab/Acne_Multimodal_Nof1/tree/main/Data}). 

In this acne pilot study, five participants conducted a series of N-of-1 trials alternating between using an acne cream (three participants used cream A, the other two participants cream B) and no intervention. The trials lasted 16 days in total, with 3 measurements (i.e. images) per day, intervention and non-intervention phases of two days each, in an alternating design.
The study resulted in a total of 255 images for all participants. The images were taken in a standardized setting with fixed rooms, lighting, and position as well as the same phone every time. Additionally, temperature at the time of taking the image, application of lotion or make-up, and a treatment indicator were registered.

Each image was scored blindly by 5 raters in terms of acne severity, which was then standardized on a scale between 0 and 1 and averaged. These average acne severity scores serve as a baseline reference for our proposed unsupervised approach and are used in all analyses reported above.

\subsubsection{Statistical Analysis} \label{Methods - 2 Statistics}

All settings for the preprocessing of the data, the autoencoder architecture and the statistical hypothesis tests were the same as described in section \ref{Methods - Simulation Study} in the analysis of the simulated data.

\section*{Data Availability}
The data generated in the simulation study is available at \url{https://www.doi.org/10.5281/zenodo.17054020}. The data of the published series of acne N-of-1 trials is available from \url{https://github.com/HIAlab/Acne_Multimodal_Nof1/tree/main/Data} . 

\section*{Code Availability}
All code used in the analyses of the simulation study and application is available from \url{https://github.com/HIAlab/multimodal_nof1_unsupervised} .

\section*{Acknowledgement}
No funding for the study is declared.
This project has been supported by the Deutsche Forschungsgemeinschaft (DFG, German Research Foundation), project number 459422098.

\section*{Author Contributions}
JS, TG and SK conceptualized the study and developed its methodology. JS carried out data simulation and all analyses. SK supervised the study. JS drafted the initial manuscript. SK and TG critically revised and edited the manuscript. All authors read and approved the final manuscript.

\section*{Competing Interests}
The authors declare no competing interests.

\bibliography{bibliography}
\bibliographystyle{plain}

\end{document}